\documentclass[prd,nofootinbib,showpacs,preprint]{revtex4}
\usepackage{amsmath}
\usepackage{graphicx}
\usepackage{hyperref}  
\usepackage{color} 
\usepackage{epsfig}
\usepackage{epsf}

\begin{document}
\title{Abelian Extension of Standard Model with Four Generations \\}
\author{Debasish Borah}
\email{debasish@phy.iitb.ac.in}
\affiliation{Indian Institute of Technology Bombay, Mumbai - 400076, India}

\begin{abstract}
An abelian gauge extension of the Standard Model is proposed with a fourth generation. The fourth generation fermions obtain their masses from a heavier Higgs doublet which makes no tree level contributions to the first three generations masses. Light first three generations neutrino masses continue to have type I seesaw 
explanation whereas the fourth generation neutrino turns out to be a heavy Dirac neutrino. In the minimal version of such a model with no off-diagonal Yukawa couplings between the fourth and the first three generation, such a heavy Dirac neutrino can be long lived on cosmological time scales. In this model the stated LHC exclusion range
 $120 \; \text{GeV} < m_H < 600 \; \text{GeV}$ on the lighter Higgs placed in the context of generic fourth generation standard model is evaded. Also, the Dirac 
fourth generation neutrino in this model, if stable would constitute upto $1 \%$ of the cold Dark Matter in the Universe. 
\end{abstract}

\pacs{12.60.Fr,12.60.-i,14.60.Pq}
\maketitle

\section{Introduction}
\indent The Standard model of particle physics has been phenomenologically the most successful low energy effective theory for the last few decades. The 
predictions of standard model have been verified experimentally 
with a very high accuracy except the missing Higgs boson. Despite its phenomenological succcess, we all now know that this model neither address 
many theoretical issues like 
gauge hiererchy problem, nor provides a complete understanding of various observed phenomena like non-zero neutrino masses, dark matter etc. 
A great deal of works have been done so far on various possible extensions of the Standard model, although non of them can be called a complete 
phenomenological model. Such extensions generally involve incorporating some extra symmetries into the Standard model. These symmetries may be 
an extra gauge symmetry like in left-right symmetric model \cite{Pati:1974yy,
Mohapatra:1974gc, Senjanovic:1975rk, Mohapatra:1980qe, Deshpande:1990ip}, Grand Unified Theories (GUT) \cite{Georgi:1974sy} etc. 
Another highly motivating symmetry is supersymmetry, the symmetry between bosons and fermions. The inclusion of supersymmetry into the 
Standard model (MSSM) has many advantages among which stabilizing the higgs mass against the radiative corrections, providing a cold dark 
matter candidate (which can be made stable by incorporating R-pariy), making the gauge coupling constants unify at high energy are significantly 
important.

\indent One very non-conventional extension of the Standard Model is to go beyond three generations of quarks and 
leptons \cite{Barger:1984jc, Frampton:1999xi,Holdom:2006mr}. Although the number of light neutrinos are constrained to three 
from Big Bang Nucleosynthesis(BBN) as well as precision measurement of Z boson decay width, there is absolutely nothing which prevents 
us from adding a heavy fourth generation with the corresponding fourth neutrino heavier than $M_Z/2= 45 \text{GeV}$. Similar lower bounds 
will exists for charged fermions also \cite{0954-3899-37-7A-075021}. We know that the smallness of three Standard Model
neutrino masses \cite{Fukuda:2001nk, Ahmad:2002jz, Ahmad:2002ka, Bahcall:2004mz} can be naturally explained 
via see-saw mechanism \cite{Minkowski:1977sc, GellMann:1980vs, Yanagida:1979as, Mohapatra:1979ia}. After incorporating a fourth 
generation neutrino into the Standard Model (SM4), the seesaw mechanism should be such that it gives three light and one very heavy neutrinos. 
Such analysis within the context of SM4 were done in \cite{Hill:1989vn}. Motivated by the idea of introducing a separate Higgs with larger vacuum expectation value (vev) to account for the heavier fermion masses such as top quark mass \cite{Das:1995df} or fourth generation fermion masses \cite{BarShalom:2011zj}, here also we propose a model with an extended Higgs sector and a fourth chiral family of quarks and leptons. However, our model differs from the earlier models in the sense that we incorporate an additional abelian gauge sector which couples to the first three generations differently than it does to the fourth generation fermions. Such non-universal gauge couplings automatically forces one to have at least two different Higgs doublets to give masses to the fermions. In our model, Majorana neutrino masses of first three generation arise after spontaneous gauge symmetry breaking whereas the fourth generation neutrino turns out to be a Dirac neutrino. We also point out that our model reproduces the Higgs-fermion structure considered by the authors in \cite{He:2011ti}. Due to the existence of a heavy Higgs doublet
which couples only to the fourth generation fermions and a lighter Higgs which couples only to the first three generations, our model can evade
the LHC exclusion range $120 \; \text{GeV} < m_H < 600 \; \text{GeV}$ placed within the context of generic SM4 \cite{Koryton2011}. The fourth generation Dirac neutrino can be long lived in this minimal model if we set the off diagonal Yukawa couplings between first three generations and and the fourth generation to zero. We further show that the a long lived heavy Dirac
fourth generation neutrino can contribute upto $1 \%$ of the total dark matter in the Universe.

Fourth generation chiral fermions can have many other interesting phenomenological 
consequences, for example in rare B and K decays \cite{Soni:2010xh}. It can also account for like-sign dimuon charge asymmetry observed by 
D0 recently \cite{Abazov:2010hv} as was discussed in \cite{Choudhury:2010ya}. However here we restrict ourselves to the issue of neutrino mass 
and dark matter only. 

\indent This paper is organized as follows. In the next section \ref{sec1} we discuss the  $U(1)_X$ extended Standard Model with four generations, 
discuss the spontaneous gauge symmetry breaking and neutrino mass. We briefly comment on the fourth generation and LHC Higgs search in 
section \ref{4thlhc}. In section \ref{4thdm} we calculate the relic abundance of a stable fourth generation Dirac neutrino within 
the $U(1)_X$ model framework and then conclude in section \ref{sec3}. 

\section{$U(1)_X$ extended model with purely Dirac Fourth Generation Neutrino}
\label{sec1}
Abelian gauge extension of Standard Model is one of the best motivating examples of beyond Standard Model physics. 
For a review see \cite{Langacker:2008yv}. Such a model is also motivated within the framework of GUT models, for example $E_6$. 
The supersymmetric version of such models have an additional advantage in the sense that they provide a solution to the MSSM $\mu$ problem. 
Here we consider an extension of the Standard Model gauge group with one abelian $U(1)_X$ gauge symmetry. Thus the model we are going to 
work on is an $SU(3)_c \times SU(2)_L \times U(1)_Y \times U(1)_X$ gauge theory with four chiral generations. We will consider family 
non universal $U(1)_X$ couplings such that the first three generations and the fourth generation have different charges under $U(1)_X$. 
Since the coupling is universal in the first three generations and we set the off diagonal Yukawa couplings between first three generation and the fourth generation to zero in this minimal model there will not be any severe constraints from Flavor Changing Neutral Current 
(FCNC) limits. As it will be clear later, our purpose of choosing such non universal couplings is to allow Majorana mass terms for the first 
three generation right handed neutrinos only and not for the fourth generation right handed neutrino. 

\subsection{The Matter Content}

The fermion content of our model is 
\begin{equation}
Q_i=
\left(\begin{array}{c}
\ u \\
\ d
\end{array}\right)
\sim (3,2,\frac{1}{6},n_1),\hspace*{0.8cm}
L_i=
\left(\begin{array}{c}
\ \nu \\
\ e
\end{array}\right)
\sim (1,2,-\frac{1}{2},n_4), \nonumber 
\end{equation}
\begin{equation}
u^c_i \sim (3^*,1,\frac{2}{3},n_2), \quad d^c_i \sim (3^*,1,-\frac{1}{3},n_3), \quad e^c_i \sim (1,1,-1,n_5), \quad \nu^c_i \sim (1,1,0,n_6) \nonumber 
\end{equation}
where $ i=1,2,3 $ goes over the three generations of Standard Model and the numbers in the bracket correspond to the quantum number under the gauge group $SU(3)_c \times SU(2)_L \times U(1)_Y \times U(1)_X$. Similarly the fourth generation fermions are 
\begin{equation}
Q_4=
\left(\begin{array}{c}
\ u \\
\ d
\end{array}\right)
\sim (3,2,\frac{1}{6},n_8),\hspace*{0.8cm}
L_4=
\left(\begin{array}{c}
\ \nu \\
\ e
\end{array}\right)
\sim (1,2,-\frac{1}{2},n_{11}), \nonumber 
\end{equation}
\begin{equation}
u^c_4 \sim (3^*,1,\frac{2}{3},n_{10}), \quad d^c_4 \sim (3^*,1,-\frac{1}{3},n_9), \quad e^c_4 \sim (1,1,-1,n_{12}), \quad \nu^c_4 \sim (1,1,0,n_{13}) \nonumber 
\end{equation}
The quantum numbers of the fermions under the new gauge symmetry $U(1)_X$ should satisfy the following anomaly cancellation conditions:
$$ [SU(3)_c]^2U(1)_X: 3(2n_1-n_2-n_3)+(2n_8-n_9-n_{10}) = 0$$
$$ [SU(2)_L]^2U(1)_X: 3(\frac{3n_1}{2}+\frac{n_4}{2})+(\frac{3n_8}{2}+\frac{n_{11}}{2})=0 $$
$$ [U(1)_Y]^2U(1)_X: 3(\frac{n_1}{6}-\frac{4n_2}{3}-\frac{n_3}{3}+\frac{n_4}{2}-n_5)+(\frac{n_8}{6}-\frac{4n_{10}}{3}-\frac{n_9}{3}+\frac{n_{11}}{2}-n_{12}) $$
$$U(1)_Y[U(1)_X]^2: 3(n^2_1-2n^2_2+n^2_3-n^2_4+n^2_5)+(n^2_8-2n^2_{10}+n^2_9-n^2_{11}+n^2_{12})=0$$
$$[U(1)_X]^3: 3(6n^3_1-3n^3_2-3n^3_3+2n^3_4-n^3_5)
+(6n^3_8-3n^3_{10}-3n^3_9+2n^3_{11}-n^3_{12})-3n^3_6-n^3_{13}=0$$
$$U(1)_X: 3(6n_1-3n_2-3n_3+2n_4-n_5)+(6n_8-3n_9-3n_{10}+2n_{11}-n_{12})-3n_6-n_{13} = 0$$
We consider one possible solution to the above anomaly matching conditions which allows us to choose a minimal scalar sector for our purposes:
$$n_1=n_4=n_8=n_{11}=0, \quad n_2=-n_3 = -n_5=n_6,\quad n_9 = -n_{10}=n_{12}=-n_{13}$$ 
Here we are interested in a solution where $ n_2 \neq n_{9} $.
\subsection{Fermion and Gauge Boson Masses}
For the particular solution of the anomaly matching 
conditions mentioned in the previous subsection, the Higgs fields required to give rise to the fermion Dirac masses are $H_1 (1,2,-\frac{1}{2},n_2) $ and $ H_2 (1,2,-\frac{1}{2},n_{10})$ 
where $H_1$ gives rise to the first three generation Dirac masses and the latter gives rise to the fourth generation Dirac masses. Since the Higgs 
scalars do not contribute to the anomalies we can choose their representation under the gauge group independent of the above anomaly matching conditions. 
Since $ n_{13} \neq 0 $, it is clear that the Majorana mass term for the fourth generation singlet neutrino is not allowed in the Lagrangian although 
the same is allowed for the first three generations if we include one more Higgs $S(1,1,0,-2n_2)$ whose vev will give rise to the Majorana mass term of 
the first three generation neutrinos. This singlet Higgs field also plays a crucial role in the gauge symmetry breaking resulting in a heavy $U(1)_X$ 
boson. It also keeps the scale of $U(1)_X$ symmetry breaking higher than the electroweak symmetry breaking. Due to the absence of Majorana mass term of 
fourth generation neutrino, we arrive at three light ($ \sim \text{eV}$) standard model Majorana neutrinos and one Dirac neutrino. This is exactly what 
we want: a fourth generation Dirac neutrino whose mass can be easily adjusted to lie above the experimental lower bounds.
The Yukawa Lagrangian is 
$$\mathcal{L}_Y = Y_u \overline{Q}_L H_1 u_R + Y_d \overline{Q}_L H^{\dagger}_1d_R+Y_{\nu} \overline{L} H_1 N_R+ Y_e \overline{L}H^{\dagger}_1 e_R+f S N_R N_R $$
$$ +Y^{(4)}_u \overline{Q^{(4)}}_L H_2 u^{(4)}_R + Y^{(4)}_d \overline{Q^{(4)}}_L H^{\dagger}_2d^{(4)}_R+Y^{(4)}_{\nu} \overline{L^{(4)}} H_2 N^{(4)}_R+ Y^{(4)}_e \overline{L^{(4)}}H^{\dagger}_2 e^{(4)}_R$$
Let $\langle H_{1,2} \rangle = v_{1,2}, \langle S \rangle = s$. For the minimal version of our model, we set the off diagonal Yukawa couplings between first three generations and the fourth generation to zero that is, $Y_{4i} = 0$ where $i = 1,2,3$. Thus the first three generation right handed neutrino acquire Majorana neutrino mass proportional to $ f \langle S \rangle = fs $ whereas the fourth generation right handed neutrino does not acquire any Majorana mass.
Hence the first three generation neutrino mass arise from seesaw mechanism whereas the fourth generation neutrino is a Dirac neutrino due to the absence of corresponding Majorana mass term. Under the assumption that the off diagonal Yukawa couplings between first three and the fourth generation are zero, the fourth generation neutrino can be a long lived particle and hence can play a non-trivial role in cosmology. We pursue this study in the next section in the context of dark matter. It may be noted that, choosing a non-minimal
solution to the anomaly matching conditions and hence an extended Higgs sector, the off diagonal Yukawa couplings can naturally be set to zero.

The gauge boson masses will come from the kinetic terms of the Higgs fields. Denoting the $SU(2)_L, U(1)_Y, U(1)_X$ gauge fields as $(W^{\mu}_1, W^{\mu}_2, W^{\mu}_3), Y^{\mu}, X^{\mu}$ respectively, we write the neutral gauge boson mass matrix in the $(W^{\mu}_3, Y^{\mu}, X^{\mu})$ basis as
\begin{equation}
M =\frac{1}{2}
\left(\begin{array}{cccc}
\ g^2_2(v^2_1+v^2_2) & g_1g_2(v^2_1+v^2_2) &  -g_2g_x(n_2v^2_1+n_{10}v^2_2) \\
\ g_1g_2(v^2_1+v^2_2) & g^2_1(v^2_1+v^2_2) & -g_1g_x(n_2v^2_1+n_{10}v^2_2) \\
\ -g_2g_x(n_2v^2_1+n_{10}v^2_2) & -g_1g_x(n_2v^2_1+n_{10}v^2_2)  & 4g^2_x(n^2_2v^2_1+n^2_{10}v^2_2+4n^2_2s^2)
\end{array}\right)
\end{equation}
The off-diagonal elements of the above mass matrix indicate non-zero mixings between the Standard Model gauge bosons and the $U(1)_X$ boson. The non-zero mixing arise since the doublet Higgs fields $H_1, H_2$ are charged under both the Standard Model gauge group as well as the extra $U(1)_X$. However these mixing have to be very small so as not to be in conflict with the electroweak precision measurements. The simplest way to evade all these restrictions is to consider zero mixings which can be achieved simply by the following constraint:
$$ n_2v^2_1+n_{10}v^2_2 = 0 $$
The charged $W$ boson mass is $M^2_W = \frac{1}{2}g^2_2(v^2_1+v^2_2)$. Using this and the above constraint we get
\begin{equation}
v^2_1 = \frac{2 M^2_W}{g^2_2}\frac{n_{10}}{n_{10}-n_2}
\end{equation}
\begin{equation}
v^2_2 = \frac{2 M^2_W}{g^2_2}\frac{-n_2}{n_{10}-n_2}
\end{equation}
Here we restrict our discussion to this simple situation of no-mixing with the extra $U(1)_X$ boson. Thus the first three generation neutrino mass comes from the usual Type I seesaw formula
$$ m_{\nu} = -\frac{v^2_1}{M_R} Y_{\nu}\, Y^T_{\nu} $$
And the fourth generation neutrino has a Dirac mass $ m_{\nu4} = Y^{(4)}_{\nu} v_2 $. The lower bounds on the fourth generation charged fermion masses \cite{0954-3899-37-7A-075021} are 
$$ m_{t'} \geq 256 \text{GeV}, \quad m_{b'} \geq 199 \text{GeV}, \quad m_{\tau'} \geq 100 \text{GeV}$$
which can be satisfied by suitable choice of Yukawa couplings. The requirement of perturbativity of Yukawa couplings $(Y^2 < 4\pi)$ at the electroweak scale 
restricts the vev of $H_2$ to be at most $256/\sqrt{4 \pi} $ and hence
\begin{equation}
\frac{-n_2}{n_{10}-n_2} > \frac{g^2_2}{2 M^2_W} \frac{256}{\sqrt{4 \pi}} 
\end{equation}
In this particular model, we assume that the fourth generation fermions have no mixing with the first three generations at tree level and hence the fourth generation neutrino (if lighter than the corresponding charged fermion) can be stable or long-lived and hence may play a role as Dark Matter. We do the analysis of relic density of such a stable heavy neutrino in the next section.

\section{Fourth Generation and Higgs searches at LHC}
\label{4thlhc}
For the last one year, LHC has been producing large number of data sets to confirm many of the already established facts as well as 
to rule out a lot of parameter space for many beyond standard model frameworks. For standard model with four generations, CMS has ruled out 
the Higgs boson mass in the range of $\sim 120-600 \; \text{GeV}$ at $95\%$ C.L \cite{Koryton2011}.

Some considerations on the implications of a fourth generation that may
evade this bound have appeared recently in \cite{He:2011ti}. As pointed out in \cite{He:2011ti}, it is possible to reduce the tension between Higgs searches at the LHC and a heavy fourth generation 
by extending the scalar sector of the standard model. The LHC exclusion range $120 \; \text{GeV} < m_H < 600 \; \text{GeV}$ \cite{Koryton2011} corresponds to the process $gg \rightarrow H \rightarrow VV $ where $V$ can be either $W^{\pm}$ or $Z$ bosons. The authors of \cite{He:2011ti} considered two Higgs doublets $H_1, H_2$ with the first 
one coupling to the first three generations and the second one coupling to the fourth generation only. Choosing the Higgs mixing angle $\alpha$ to be same as $ \beta = \tan^{-1}{v_2/v_1}$ makes the coupling of one of the Higgs mass eigenstates to the vector boson zero and hence can be light without conflicting with the LHC exclusion range coming from the $gg \rightarrow H \rightarrow VV $ process. The other Higgs which couple to vector bosons can be as heavy as the generic unitarity bound for two Higgs 
doublet models $\sim 700 \; \text{GeV}$ \cite{Casalbuoni:1987cz,Kanemura:1993hm,Akeroyd:2000wc,Horejsi:2005da}, such a Higgs can still be 
outside the LHC search range. 

Our model has a different motivation and differs from \cite{He:2011ti} in that the extra $U(1)_X$ gauge sector prevents bilinear mixings of the Higgs doublets introduced. The dimensionless coupling of the quartic mixing terms can be chosen to be small enough so that the lightest neutral Higgs mass eigenstate has negligible coupling to the fourth generation fermions. Thus, although both of them couple to vector bosons, the lighter Higgs have no coupling to the fourth generation fermions and hence can evade the LHC exclusion range coming from $gg \rightarrow H \rightarrow VV $ process. The heavier Higgs couple to fourth generation only and can be as heavy as the unitarity bound to stay outside the current LHC search range. It may be noted that in \cite{He:2011ti} there is no physical mechanism to prevent the extra Higgs doublet vev from feeding into the lighter fermion masses whereas our model provides an explanation for this by the non-universal couplings 
of the extra $U(1)_X$ gauge boson to the fermions which allows only $H_1$-first three generation and $H_2$-fourth generation couplings.

\section{Fourth Generation Dirac Neutrino as Dark Matter}
\label{4thdm}
The relic abundance of a dark matter particle $\chi$ is given by the the Boltzmann equation
\begin{equation}
\frac{dn_{\chi}}{dt}+3Hn_{\chi} = -\langle \sigma v \rangle (n^2_{\chi} -(n^{eqb}_{\chi})^2)
\end{equation}
where $n_{\chi}$ is the number density of the dark matter particle $\chi$ and $n^{eqb}_{\chi}$ is the number density when $\chi$ was in thermal equilibrium. $H$ is the Hubble rate and $ \langle \sigma v \rangle $ is the thermally averaged annihilation cross section of the dark matter particle $\chi$. In terms of partial wave expansion $ \langle \sigma v \rangle = a +b v^2$. Numerical solution of the Boltzmann equation above gives \cite{Kolb:1990vq}
\begin{equation}
\Omega_{\chi} h^2 \approx \frac{1.04 \times 10^9 x_F}{M_{Pl} \sqrt{g_*} (a+3b/x_F)}
\end{equation}
where $x_F = m_{\chi}/T_F$, $T_F$ is the freeze-out temperature, $g_*$ is the number of relativistic degrees of freedom at the time of freeze-out. Dark matter particles with electroweak scale mass and couplings freeze out at temperatures approximately in the range $x_F \approx 20-30$. This again simplifies to \cite{Jungman:1995df}
\begin{equation}
\Omega_{\chi} h^2 \approx \frac{3 \times 10^{-27} cm^3 s^{-1}}{\langle \sigma v \rangle}
\end{equation}
The thermal averaged annihilation cross section $\langle \sigma v \rangle$ is given by \cite{Gondolo:1990dk}
\begin{equation}
\langle \sigma v \rangle = \frac{1}{8m^4T K^2_2(m/T)} \int^{\infty}_{4m^2}\sigma (s-4m^2)\surd{s}K_1(\surd{s}/T) ds
\end{equation}
where $K_i$'s are modified Bessel functions of order $i$, $m$ is the mass of Dark Matter particle and $T$ is the temperature. We consider two annihilation cross sections of cosmological importance, namely $\sigma ( \nu_4 \bar{\nu}_4 \rightarrow f \bar{f}) $ and  $\sigma ( \nu_4 \bar{\nu}_4 \rightarrow W^+ W^-) $ where $f$ is any Standard Model fermion. The first process can take place via s-channel exchange of $Z$ and $X$ boson. The second process can take place via s-channel $Z$ or Higgs boson $H$ exchange or t-channel exchange of charged leptons. Since there is no mixing of fourth generation with the first three generation, this charged lepton is the fourth charged lepton $\tau'$. In our model, we always consider $m_{\tau'} > m_{\nu4}$ so as to make the fourth generation neutrino perfectly stable. Thus only the s-channel annihilation processes are of interest. The cross section for $f\bar{f}$ final state through $Z$ boson exchange and $W^+ W^-$ final has been calculated in \cite{Enqvist:1988we} and we use their standard result. The s-channel $X$ boson exchange cross section is 
\begin{equation}
\sigma_X(\nu_4 \bar{\nu}_4 \rightarrow f \bar{f}) = \frac{N_c g^4_x}{32\pi s}\frac{\beta_f}{\beta_{\nu4}} \lvert D_x \rvert^2 [ (v^2_f+a^2_f)(v^2_{\nu4}+a^2_{\nu4})\frac{s^2}{4}(1+\frac{\beta^2}{3})+(v^2_f-a^2_f)(v^2_{\nu4}+a^2_{\nu4})m^2_f(s-2m^2_{\nu4}) \nonumber \\
\end{equation}
\begin{equation}
+(v^2_f+a^2_f)(v^2_{\nu4}-a^2_{\nu4})m^2_{\nu4}(s-2m^2_f)+4(v^2_f-a^2_f)(v^2_{\nu4}-a^2_{\nu4})m^2_f m^2_{\nu4} ]
\end{equation}
where $v$,$a$ are the vector and axial couplings of fermions to the $X$ boson respectively; $N_c$ is the color factor which is 3 for quarks and 1 for leptons, $\lvert D_x \rvert^2 = 1/[(s-M^2_x)^2+ \Gamma^2_xM^2_x] $ and $\beta$'s are defined as 
$$ \beta_f = \Big(1-\frac{4m^2_f}{s}\Big)^{1/2}, \quad \beta_{\nu4} = \Big(1-\frac{4m^2_{\nu4}}{s}\Big)^{1/2}, \quad \beta = \beta_f \beta_{\nu4} $$
\begin{figure}[htb]
\centering
\includegraphics{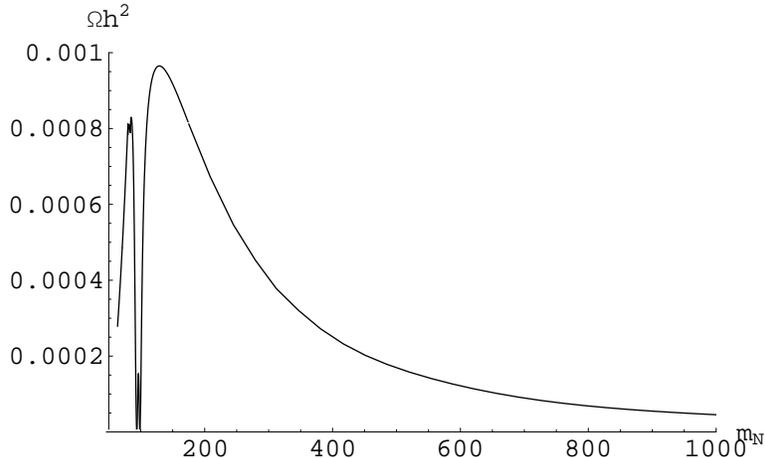}
\caption{Relic abundance of fourth generation Dirac Neutrino as a function of its mass}
\label{fig1}
\end{figure}
The relic density as a function of the fourth generation neutrino mass $m_N$ is shown in the figure \ref{fig1}. In this mass range $M_Z/2 < m_N < M_W$, the only possible annihilation channels are the $N \bar{N} \rightarrow f \bar{f}$ through s-channel exchange of Z-boson or X-boson(depending on the mass of $M_X$). In this particular example, the $U(1)_X$ gauge charges are chosen as $n_2 =-1, n_{10} = \frac{1}{2}$. Also we have taken the 
extra $U(1)_X$ coupling to be $10^{-2}$ and the $U(1)_X$ symmetry breaking scale to be $5$ TeV which results in $M_X = 142 \; \text{GeV}$. Thus the X-boson mediating channel opens only when $m_N \geq 71 \; \text{GeV} $. As we go beyond this mass range, more and more annihilation channels become important and hence reduces the relic abundance further. It is seen from the figure \ref{fig1} that
the fourth generation Dirac neutrino can at most give rise to $1 \%$ of the total dark matter abundance estimated by Seven-Year Wilkinson Microwave Anisotropy Probe (WMAP) Observations \cite{Jarosik:2010iu} 
\begin{equation}
\Omega_{DM}h^2 = 0.1123 \pm 0.0035
\end{equation} 
Thus fourth generation Dirac neutrino if stable or long lived, like in 
the minimal version of our model, can give rise to a very small fraction of total dark matter in the Universe and hence of little cosmological
 significance. However it can have distinct collider signatures. Being heavy and stable, it can give rise to a large missing transverse 
 energy in the colliders.
 
\section{Conclusion}
\label{sec3}
We have studied one possible framework which gives rise to the experimentally allowed neutrino mass spectra in four generation models. The Standard Model gauge group is enhanced to include one extra abelian gauge symmetry $U(1)_X$ under which the fourth generation fermions transform differently from the first three generations. We have assumed zero mixing between the fourth and first three generations and showed that there can be a common seesaw
 mechanism which can generate three light standard model neutrinos and one heavy stable fourth generation Dirac neutrino. We also point out that 
our model reproduces the Higgs-fermion structure considered by the authors of \cite{He:2011ti} with certain differences. Gauge structure of our model naturally prevents bilinear mixing between the two Higgs doublets and provides a physical mechanism why one doublet couples only to the first three generations and the other couples only to the fourth generation. Thus the Higgs which couple to the fourth generation only can be as heavy as the unitarity bound so as to evade the current LHC exclusion range for generic
standard model with four generations.The lighter Higgs do not couple to fourth generation at tree level and hence can also evade the LHC exclusion range.

We also consider the possibility of such a heavy neutrino as dark matter and found out the relic density. However, we find that for the mass range $M_Z/2 < m_{\nu4} <1 \; \text{TeV}$, fourth generation Dirac neutrino can at most give rise to $1\%$ of the total dark matter in the Universe and hence can play a role in multi component dark matter formalisms. It can also have important collider signatures in terms of missing transverse energies the details of which we have skipped in our present work.

\section{Acknowledgement}
This work was partially supported by a grant from Department of Science and Technology (DST), Govt. of India. The author would like to thank Prof Urjit A. 
Yajnik, IIT Bombay for useful comments and discussions.
\bibliographystyle{apsrev}

\end{document}